\documentclass{article} 
\let\counterwithout\relax 
\usepackage{arxiv}
\usepackage[latin1,utf8]{inputenc}
\usepackage[english]{babel}
\usepackage[square,sort,comma,numbers]{natbib}
\usepackage{lmodern}
\usepackage{amssymb}
\usepackage{amsmath}
\usepackage{mathtools}
\usepackage{graphicx}
\usepackage{cuted}
\usepackage[usenames,dvipsnames]{xcolor}
\usepackage{tikz}
\usepackage{hyperref}
\usepackage{upgreek}
\usepackage{glossaries}
\usepackage{bm}
\usepackage{algorithm}
\usepackage[noend]{algpseudocode}
\usepackage[caption=false]{subfig}
\usepackage{alphalph}
\usepackage{chngcntr}
\usepackage{dashrule}
\usepackage{footnote}
\usepackage{footmisc}
\usepackage{nicefrac}       
\usepackage{microtype}      
\usetikzlibrary{decorations.text,shapes,arrows,shapes.multipart,calc,positioning,matrix,backgrounds,folding,
 decorations.pathmorphing,spy,fit,decorations.fractals,decorations.markings,external}
\PassOptionsToPackage{hyphens}{url}\usepackage{hyperref}

\hypersetup{%
        pdfborder={0 0 0},
        colorlinks= true,
        linkcolor= blue,
        citecolor= blue,
        urlcolor= black,
        pdfmenubar= true,
        pdfnewwindow = true,
        pdfencoding  = unicode,
        pdfauthor={Juan F. Restrepo},
        pdftitle={Transfer entropy rate through Lempel-Ziv's complexity}
}
\counterwithout{figure}{section}

\graphicspath{{./figs/}{./figs/tikz_img/}}
\DeclareGraphicsExtensions{.eps,.pdf}
\makeatletter
\newcommand{\algrule}[1][.2pt]{\par\vskip.5\baselineskip\hrule height #1\par\vskip.5\baselineskip}
\makeatother

\newacronym{snr}{SNR}{Signal to Noise Ratio}
\newacronym{etal}{et al.}{and others}
\newacronym{pdf}{pdf}{probability density function}
\newacronym{lzc}{LZC}{Lempel-Ziv's complexity}
\newacronym{te}{TE}{Transfer Entropy}
\newacronym{ter}{TER}{Transfer Entropy Rate}
\newacronym{lzte}{LeZTER}{Lempel-Ziv based Transfer Entropy Rate}
\newacronym{knn}{KNN}{K-nearest-neighbor}
\newcommand{\E}[1]{\left(#1\right)}                                             
\newcommand{\Q}[1]{\left[#1\right]}                                             
\newcommand{\W}[1]{\left\{#1\right\}}                                           
\newcommand{\Lim}[3]{\mathop{\lim}                                              
                     \limits_{#1\to#2}\!{#3}}
\newcommand{\Tyx}[2]{\text{\textbf{T}}_{#1\to#2}^{\E{m}}}                      
\newcommand{\xv}[1]{\bm{x}_{#1}^{\E{m}}}                                        
\newcommand{\Xv}[1]{{X}_{#1}^{\E{m}}}                                        
\newcommand{\Yv}[1]{{Y}_{#1}^{\E{m}}}                                        
\newcommand{\Yvs}[1]{{\hat{Y}}_{#1}^{\E{m}}}                                 
\newcommand{\Vv}[1]{\bm{V}_{#1}}                                                
\newcommand{\Cc}{\mathcal{C}}                                                   
\newcommand{\MatV}{\mathcal{V}}                                                 
\newcommand{\tyx}[2]{\text{\textbf{t}}_{#1\to #2}^{\E{m}}}                      
\newcommand{\tyxs}[2]{\text{\textbf{\^{t}}}_{\widehat{#1}\to #2}^{\E{m}}}       
\newcommand{\TotalTER}{\mathcal{T}}                                             

\begin{document}
\title{Transfer Entropy Rate\\Through Lempel-Ziv Complexity}
\author{%
  Juan~F.~Restrepo\\
    Laboratorio de Señales y Dinámicas  no Lineales\\
    Instituto de Investigación y Desarrollo en Bioingeniería y Bioinformática\\
    CONICET - Universidad Nacional de Entre Ríos.\\
    Ruta prov  11 km 10 Oro Verde, Entre Ríos, Argentina.\\
    \texttt{jrestrepo@ingenieria.uner.edu.ar}\\
\And%
  Diego~M.~Mateos\\
   Instituto de  Matemática Aplicada del Litoral - CONICET - UNL.\\
   CCT CONICET, Santa Fe, Colectora Ruta Nac.  N 168, Paraje El Pozo, 3000 Santa Fe, Argentina.\\
   Facultad de Ciencia  y  Tecnología.  Universidad  Autónoma de  Entre  Ríos.\\
   Ruta  prov 11  km  10  Oro  Verde  -  Entre  Ríos - Argentina\\
   \texttt{mateosdiego@gmail.com}\\
\And%
  Gastón~Schlotthauer\\
    Laboratorio de Señales y Dinámicas  no Lineales\\
    Instituto de Investigación y Desarrollo en Bioingeniería y Bioinformática\\
    CONICET - Universidad Nacional de Entre Ríos.\\
    Ruta prov  11 km 10 Oro Verde, Entre Ríos, Argentina.\\
    \texttt{gschlotthauer@conicet.gov.ar}
    \thanks{This  work  was supported  by  the National  Scientific  and  Technical  Research  Council  (CONICET) of
    Argentina,   the  National  University   of  Entre  Ríos  (UNER),   and   the  Grants:   PID-6171  (UNER),   and
    PIO-14620140100014CO (UNER-CONICET).}
}
\maketitle
\begin{abstract}
  In this article we present a methodology to  estimate the \acrlong{ter} between two systems through the Lempel-Ziv
  complexity.  This methodology  carries a  set of practical  advantages:  it estimates  the \acrlong{ter}  from two
  single discrete series of measures,  it is not computationally expensive  and it does not assume any model for the
  data.  The  results  of  simulations over  three  different  unidirectional  coupled  systems,  suggest  that this
  methodology can be used to assess the direction and strength of the information flow between systems.
\end{abstract}
\keywords{Transfer Entropy, Transfer Entropy Rate, Lempel-Ziv Complexity.}
\section{\label{intro}Introduction}
The \gls{te}  and the \gls{ter} are  closely related concepts  that measure  information transport.  The  former was
proposed by  Schreiber in~\cite{Schreiber2000}  and independently by  Palu\v{s} in~\cite{Palus2001}.  The  later was
described by Amblard~\glsname{etal}  in~\cite{Amblard2009,Amblard2011}.  They are able to quantify  the strength and
direction of the coupling between simultaneously observed systems~\cite{Kaiser2002}.  Moreover,  they have become of
general   interest  since   they  can   be  used   to   study   complex   interaction   phenomena   found   in  many
disciplines~\cite{Bossomaier2016}.

On the other hand,  \gls{lzc} is a classical measure that,  for ergodic sources,  relates the concepts of complexity
(in the Kolmogorov-Chaitin sense),  and entropy rate~\cite{Zozor2005, Blanc2011}.  For an ergodic dynamical process,
the amount of new information  gained per unit of time (entropy rate) can be  estimated by measuring the capacity of
this source to generate  new patterns (\gls{lzc}).  Because of the simplicity of  the \gls{lzc} method,  the entropy
rate  can  be  estimated   from   a   single   discrete   sequence   of   measurements   with  a  low  computational
cost~\cite{Estevez-Rams2013}.

In this article we aim to relate the concepts of Transfer Entropy Rate and Lempel-Ziv complexity.  To be precise, we
will exploit the  advantages of the \gls{lzc} methodology  to calculate the \gls{ter} between  two ergodic dynamical
systems.

The remainder  of this paper  is organized as  follows.  Section~\ref{sec:metho} begins with  a brief review  of the
concepts of \gls{te},  \gls{ter} and \gls{lzc}.  In Section 3  we described the proposed methodology to estimate the
Transfer Entropy Rate  through  the  Lempel-Ziv  complexity.  In  Section~\ref{sec:Res}  we  present and analyze the
results of the simulations carried out to evaluate the performance of our approach.  Finally, Sections~\ref{sec:Dis}
and~\ref{sec:Con} the discussion and conclusions are presented.

\section{\label{sec:metho}Methodology}
In this section we will briefly review some theoretical concepts related with the \gls{lzc}, \gls{te} and \gls{ter}.
Moreover,  we will introduce the  notation used along the document.

Since our intent is to investigate a possible causality connection between two dynamical systems, we need to analyze
the signals  that they  produce.  We will assume  the existence of  ergodic probability  measures that  describe the
density of  trajectories in phase  space,  such us it  can be treated  as probability densities.  This  allows us to
analyze the dynamics of the systems through the construction of random processes from their signals.

Consider  a  system  ${X}$  that  produces a  time  series  ${x_t=x_1\dots  x_T}$.  We  can  compose  samples  of an
$m$-dimensional  time-embedded process  $\W{X^{\E{m}}}=\W{X_1,\dots,X_m}$ by  sampling  $x_t$  with  a  frequency of
$1/\tau$~\cite{Schuster2005,Granero-Belinchon2017}:
\begin{equation*}
  \xv{n} = \E{x_n,x_{n-\tau},\dots,x_{n-\E{m-1}\tau}},
\end{equation*}
where $n=1,\dots,T-\E{m-1}\tau$. The process $\W{X^{\E{m}}}$ is characterized  by the joint  probability
distribution:
\[
  p\E{\xv{n}}=P\W{\E{X_1,\dots,X_m}=\xv{n}}.
\]

We can define the $m$-order entropy rate as~\cite{Granero-Belinchon2017}:
\begin{align*}
  h\E{X^{\E{m}}} &= H\E{X_{t+\tau}|X^{\E{m}}}\\
                 &= H\E{X^{\E{m+1}}}-H\E{X^{\E{m}}},
\end{align*}
where $H\E{X^{\E{m}}}$ is the entropy of the joint distribution $p\E{\xv{n}}$:
\begin{align*}
  H\E{X^{\E{m}}}&=H\E{X_1,\dots,X_m},\\
                &=-\sum_{x_1}\cdots\sum_{x_m}{p\E{\xv{n}}\ln p\E{\xv{n}}}.
\end{align*}
The $m$-order  entropy rate measures the  variation of the total  information in the time-embedded  process when the
embedding dimension $m$  is increased by 1.  From  this definition we can  calculate the entropy rate  of the system
${X}$ as~\cite{Papoulis1991,Cover2005}:
\begin{align}
   h\E{X} &= \Lim{m}{\infty}{h\E{X^{\E{m}}}},\label{eq:entropy_rate_01}\\
          &= \Lim{m}{\infty}{\frac{H\E{X^{\E{m}}}}{m}}\label{eq:entropy_rate_02}.
\end{align}

Equations~(\ref{eq:entropy_rate_01})  and~(\ref{eq:entropy_rate_02}) relate  two  different  interpretations  of the
entropy rate.  The  first one tells  that $h\E{X}$ is a  measure of our uncertainty  about the present  state of the
system under the  assumption that its entire past is  observed.  The second one states that the  entropy rate is the
average information gained by observing the system.  In  this respect,  systems with a higher entropy rates generate
information at a higher rate, and this make their dynamics more difficult to predict.
\subsection{\label{sub:lemp}Lempel-Ziv Complexity}

The concepts of  entropy rate and Lempel-Ziv complexity  are closely related since systems  with higher entropy rate
tend to generate more complex sequences (time series).  In  that context,  the entropy rate of an ergodic system can
be estimated  by measuring its capacity  to generate  new patterns~\cite{Estevez-Rams2013}.  Estimating  the entropy
rate of a system using  the Lempel-Ziv algorithm carries a set of practical advantages:  it  can be estimated from a
single discrete series of measures, the algorithm is fast and it does not assume any model for the data.

Suppose a stationary stochastic process $\W{X_t}$ that produces a sequence $x_{t}$ of length $T$,  where for a fixed
$t$,  the random variable  $X_t$ can take values from  an alphabet $\Omega_x$ of $\alpha$  symbols.  To estimate the
complexity of  this process we  will use the  Lempel and Ziv's  scheme proposed in  1976~\cite{Lempel1976}.  In this
approach,  a sequence $x_t$ is  parsed  into  a  number  $\Cc_{x_t}$  of  words,  by  considering  as a new word any
subsequence that has  not yet been encountered.  For  example the sequence $100110111001010001011$ is  parsed in $7$
words:  $1\cdot 0\cdot 01\cdot  101\cdot  1100\cdot  1010\cdot  001011$.  Then,  the  entropy  rate  can be computed
as~\cite{Blanc2011}:
\begin{equation}
  h\E{X} = \Lim{T}{\infty}{\frac{\Cc_{x_t}\Q{\ln\E{\alpha} + \ln{\Cc_{x_t}}}}{T}}\label{eq:lz-norm}.
\end{equation}

This approach can be  easily generalized to multivariate processes by  extending the alphabet size~\cite{Zozor2005}.
Consider    an    $m$-dimensional   stationary    process    $\W{X^{\E{m}}}$,    that    produces    the   sequences
$x_{t,i}=x_{1,i},\dots,x_{T,i}$ with ${i=1,\dots,m}$,  each  one of them from an  alphabet of $\alpha$ symbols.  Let
${z_t=z_{1},\dots,z_{T}}$ be a new sequence defined over an extended alphabet of size $\alpha^m$~\cite{Zozor2005}:
\begin{equation*}
  z_t=\sum_{i=1}^{m}{\alpha^{i-1} x_{t,i}},
\end{equation*}
then the joint  Lempel-Ziv complexity ${\Cc_{x_{t,i}}=\Cc_{z_t}}$ and  the $m$-order entropy rate  can be calculated
as~\cite{Zozor2005,Blanc2011}:
\begin{align*}
  h\E{X^{\E{m}}} &= h\E{Z},\nonumber\\
   &= \Lim{T}{\infty}{\frac{\Cc_{z_t}\Q{\ln\E{\alpha^m}+\ln{\Cc_{z_t}}}}{T}}.
\end{align*}
\subsection{\label{sub:tent}Transfer Entropy and Transfer Entropy Rate}
\begin{figure*}[t!]
  \centering
  \includegraphics[width=\textwidth]{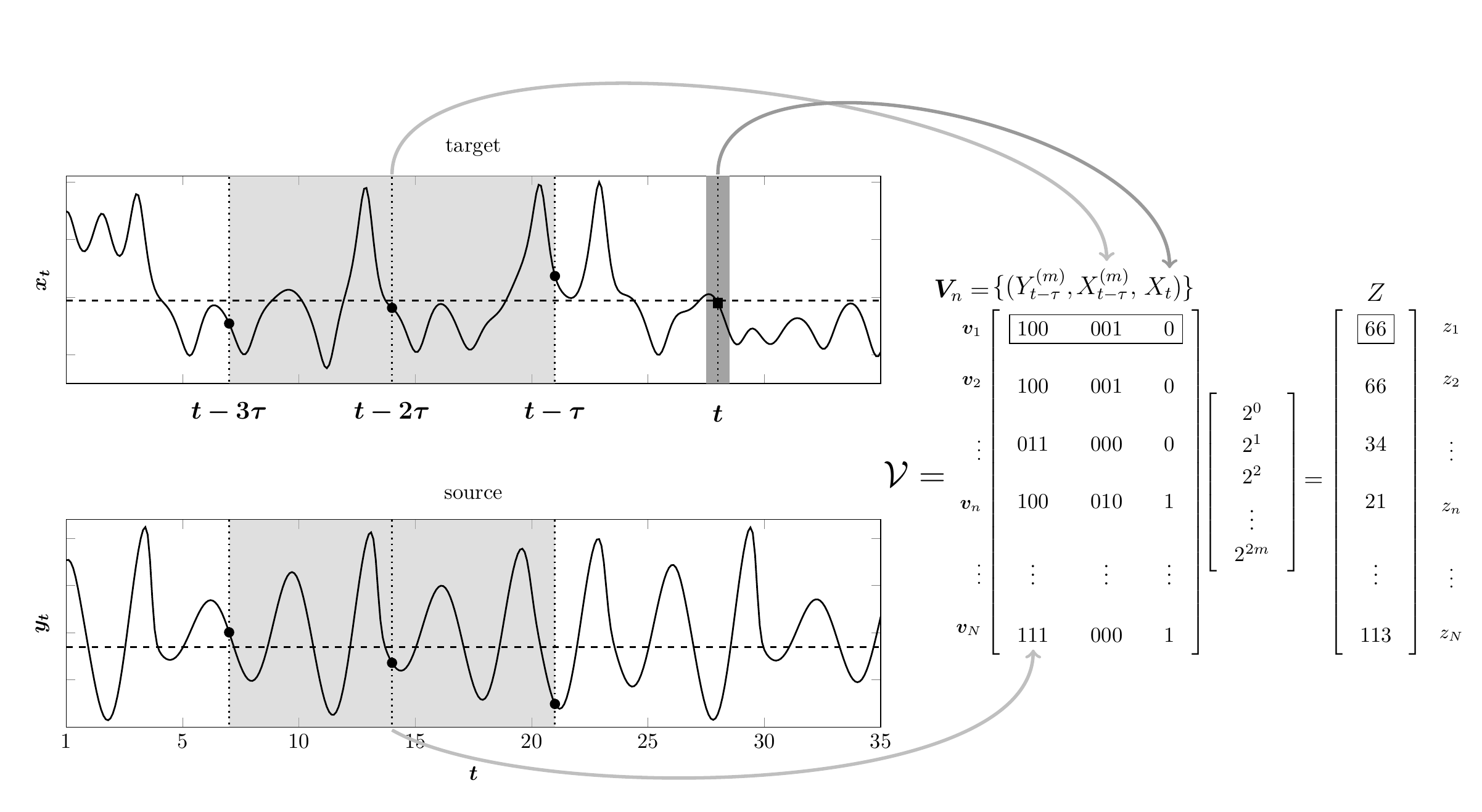}
  \caption{\label{fig:diagram}Diagram   to   obtain   the   sequence   $z_n$   to   calculate   the   entropy   rate
  $h\E{\Yv{t-\tau},\Xv{t-\tau},X_{t}}$.  The matrix $\MatV$  is obtained by embedding ($m=3$)  the binarized version
  of $x_{t}$ and $y_{t}$.  The median values of both time series are shown as horizontal dashed lines.}
\end{figure*}
The \acrlong{te} is able to  assess the amount of  information transferred from process  $Y$ (driver/source) to
process $X$ (driven/target).  It is defined as~\cite{Kaiser2002,Bossomaier2016,Palus2007,Schreiber2000}:
\begin{align}
  \Tyx{Y}{X} = &\phantom{+} H\E{\Xv{t-\tau},\Yv{t-\tau}} - H\E{X_{t},\Xv{t-\tau},\Yv{t-\tau}}, \nonumber\\
    & + H\E{X_{t},\Xv{t-\tau}} - H\E{\Xv{t-\tau}}.\label{eq:tent}
\end{align}

The parameter $m$  is commonly called the history length  or embedding dimension and $\tau$ is  the lag or embedding
lag~\cite{Bossomaier2016}.  $\Tyx{Y}{X}$ quantifies  the amount of information  contained in the  $m$-past states of
process $Y$ ($\Yv{t-\tau}$) about the current state of the process $X$ ($X_t$), that is not already explained by the
$m$-past  states  of  process $X$  ($\Xv{t-\tau}$).  This  measure  is  asymmetric  ($\Tyx{Y}{X}\neq\Tyx{X}{Y}$) and
increases  along with  the coupling  level,  allowing to  determine the  direction and  strength of  the information
flow~\cite{Kaiser2002}.

In~\cite{Amblard2009}  Amblard~\glsname{etal}  suggest  that  under  stationarity  conditions  the  \gls{te}  can be
considered    as     an    information    flow    rate.     This     idea    leads    to     the    definition    of
\acrlong{ter}~\cite{Amblard2009,Amblard2011,Haruna2013}:
\begin{align}
  \tyx{Y}{X}&  \equiv h\E{X}  - h\E{X|Y},\nonumber\\
            &= h\E{X_t,\Xv{t-\tau}}  - h\E{\Yv{t-\tau},\Xv{t-\tau},X_t},\label{eq:tent-lzc-2}
\end{align}
where $h\E{X}$ is the entropy rate of $X$ and $h\E{X|Y}$ is the conditional entropy rate~\cite{Haruna2013}:
\begin{align*}
  h\E{X|Y} &\equiv \Lim{m}{\infty}{H\E{X_t \big|\Xv{t-\tau},\Yv{t-\tau}}},\\
           &= \Lim{m}{\infty}{\frac{H\E{\Yv{t-\tau},\Xv{t-\tau},X_t}}{m}},\\
           &= h\E{\Yv{t-\tau},\Xv{t-\tau},X_t}.
\end{align*}
The \gls{ter}  lies between zero and  the entropy rate of  the target $X$,  being equal  to zero if $X$  and $Y$ are
independent~\cite{Haruna2013}.

If the processes $X$ and $Y$ had no  relationship, then $\tyx{Y}{X}$ should be equal to zero.  However,  in practical
applications,  the estimation  of $\tyx{Y}{X}$  could present a  bias due to  the finite  length of  the data.  Some
authors~\cite{Bossomaier2016} proposed to correct this bias by empirically finding the distribution of the surrogate
measurement $\tyxs{Y}{X}$.  The surrogate data must be generated in such a way that the temporal correlation between
the source and the target  is destroyed but statistical properties and the temporal  structure of both processes are
preserved~\cite{Kaiser2002,Bossomaier2016}.  Note  that  only  the  second  term  in  equation~(\ref{eq:tent-lzc-2})
depends on the source, so the surrogate transfer entropy from $Y$ to $X$ is defined as:
\begin{equation}
  \tyxs{Y}{X} = -\left<h_k\E{\Yvs{t-\tau},\Xv{t-\tau},X_t}\right>_K,\label{eq:tent-lzc-surrogate}
\end{equation}
where $\Yvs{t-\tau}$ is obtained  by redrawing with replacement samples from  $\Yv{t-\tau}$, and  $<\cdot>_K$ is the
mean value over  the  $k=1,2,\dots,  K$  surrogate  realizations.

In order to assess  the  directionality  of  the  information  transport  we  need  to  analyze the global \gls{ter}
estimator:
\begin{equation}
  \TotalTER = \tyx{Y}{X} - \tyx{X}{Y} - \E{\tyxs{Y}{X}-\tyxs{X}{Y}}.\label{eq:global_te}
\end{equation}
A positive value of $\TotalTER$ suggests that the information  flow goes from system $Y$ to system $X$,  meanwhile a
negative value suggests the contrary.  Finally, if $\TotalTER=0$, there is no information flow between systems.

\section{\label{sec:tent-lzc}Transfer Entropy  Rate Based on  Lempel-Ziv Complexity}

In this section we  formalize our approach to estimate the \acrlong{ter}  using the Lempel-Ziv complexity.  The idea
is to  estimate the two  joint entropy rates  on the right-hand  side of equation~(\ref{eq:tent-lzc-2})  by means of
their associated joint Lempel-Ziv complexities.  To this end,  we propose a methodology based on the construction of
delayed embedding  vectors from  quantized time series.  For  simplicity in the  description of  the method  we will
assume binary  time series,  although this  methodology can be  extended to higher  quantization levels\footnote{For
example,  a time  series can be  quantized into $\alpha$ symbols  using as thresholds  its own $\alpha$-quantiles.}.
Consider two  binarized time series ($\alpha=2$)  from a coupled system:  let  $x_t=x_1\dots x_T$ be  the target and
$y_t=y_1\dots y_T$ be the source.  In all the simulations, each time series was binarized with its own median value.
Set the  parameter $m$  (embedding dimension)  and $\tau$  (embedding lag)  and create  the collection  of embedding
vectors~(see~Fig.~\ref{fig:diagram}):
\begin{equation*}
  \Vv{n}=\W{\E{\Yv{t-\tau},\Xv{t-\tau}, X_t}},
\end{equation*}
where:
\begin{align*}
  n &= 1, 2,\dots, N, \qquad\text{with}\qquad N = T-m\tau,\\
  t &= m\tau + j,\\
  \bm{v}_n &= \E{\bm{y}^{\E{m}}_{t-\tau},\bm{x}^{\E{m}}_{t-\tau},x_t},\\
  \bm{x}^{\E{m}}_{t-\tau}& =\Q{x_{t-m\tau},\dots,x_{t-2\tau},x_{t-\tau}},\\
  \bm{y}^{\E{m}}_{t-\tau}& =\Q{y_{t-m\tau},\dots,y_{t-2\tau},y_{t-\tau}}.
\end{align*}

By  construction  $\Vv{n}$   is  a  collection  of   ${\E{2m+1}}\text{-uples}$  and  we  can   define  the  sequence
${z_{n}=\sum_{i=1}^{2m+1}{2^{i-1}v_{n,i}}}$,   over   an   extended   alphabet   of   size   $2^{2m+1}$  (see
Fig.~\ref{fig:diagram}).  Then, the joint entropy rate can be calculated as~\cite{Zozor2005}:
\begin{align}
  h\E{\Yv{t-\tau},\Xv{t-\tau},X_t} &= h\E{\Vv{n}},\nonumber\\
                                    &= h\E{Z},\nonumber\\
                                    &= \Lim{N}{\infty}{\frac{\Cc_{z_n}\Q{\ln\E{2^{{2m+1}}}+
                                        \ln{\Cc_{z_n}}}}{N}}\label{eq:lz-comp},
\end{align}
where $\Cc_{z_n}$ is the \gls{lzc} of the sequence $z_n$.  This procedure can be followed to estimate the first term
of      equation~(\ref{eq:tent-lzc-2}),      considering      the      collection      of      embedding     vectors
${\Vv{n}=\W{\E{\Xv{t-\tau},X_t}}}$ and  the sequence  ${z_{n}=\sum_{i=1}^{m+1}{2^{i-1}v_{n,i}}}$.  Moreover,  we can
use the  same methodology to obtain  a surrogate  measurement $h_k\E{\Yvs{t-\tau},\Xv{t-\tau},X_t}$.  In  this case,
$\Yvs{t-\tau}$  is  obtained  by  shuffling  (or  redrawing  with  replacement)  $\Yv{t-\tau}$  amongst  the  set of
$\W{\Yv{t-\tau},\Xv{t-\tau}, X_t}$ tuples.
\subsection{\label{sub:implem}Implementation}

As we have mentioned before,  our methodology is based on the construction of embedding spaces from time series.  In
this direction,  our algorithm has two parameters: the embedding dimension ($m$) and the embedding lag ($\tau$).  As
well  as  other  embedding based  algorithms,  we  have  found that  the  best  results  are  achieved  when  a good
reconstruction of  the state space  is guaranteed~\cite{Small2005}.  In other  words,  when $m$  is bigger  than the
minimum embedding dimension of the  system  and  $\tau$  is  large  enough  so  that  the various coordinates of the
embedding vectors contains as much new information as possible, without being entirely independent.  In this sense a
good choice of embedding dimension is $m=m_x + m_y+1$, here $m_x$ and $m_y$ are estimations of the minimum embedding
dimension  of  $X$  and  $Y$,  respectively.  On  the  other  hand,  we  propose  to  use  an  embedding  lag  value
$\tau=\text{max}\E{\tau_x,\tau_y}$,  where $\tau_x$ and  $\tau_y$ are the lags that  minimize the mutual information
function between $x_t$ and $x_{t-\tau}$, and between $y_{t}$ and $y_{t-\tau}$, respectively.

The Algorithm~\ref{alg:LZTE_1}  describes the  steps to calculate  the global  \acrlong{ter} ($\TotalTER$),  between
processes $X$ and $Y$.  In  the first step both time series,  $x_t=x_1\dots x_T$  and $y_{t}=y_1\dots y_T$,  must be
binarized.  This can be done using measures as the mean value,  or the median value,  among other options.  Consider
$x_{t}$ as  the target/driven series and  $y_{t}$ as the source/driver  series and  follow steps  3-6 to  obtain the
\gls{ter} from $Y$ to  $X$ ($\tyx{Y}{X}$) and step 7 to obtain its  surrogate estimation ($\tyxs{Y}{X}$).  As it was
previously described,  to estimate each  term in equation~(\ref{eq:tent-lzc-2}) we need to  embed the binarized time
series   in   spaces   with   different   dimensions.   Considering   the   one   with   the   greatest   dimension:
$\W{\Yv{t-\tau},\Xv{t-\tau},X_t}$,   the  embedding   vectors  in   this  space   can  be   disposed  in   a  matrix
${\MatV=\Q{\Yv{t-\tau},\Xv{t-\tau},X_t}}$ (see  figure~\ref{fig:diagram}).  The sequence $z_n$  can be  expressed as
the product:
\begin{equation*}
 z_n=\MatV\Q{2^0,2^1,\dots,2^{2m}}^T,
\end{equation*}
and,  the  entropy  rate  $h\E{\Yv{t-\tau},\Xv{t-\tau},X_t,}$  can  be  computed  using equation~(\ref{eq:lz-comp}).
Regarding the second  term  in  equation~(\ref{eq:tent-lzc-2}),  given  that  the  space  $\W{\Xv{t-\tau},X_t}$ is a
subspace of $\W{\Yv{t-\tau},\Xv{t-\tau},X_t}$,  the sequence $z_n$  needed to estimate $h\E{\Xv{t-\tau},X_t}$ can be
computed by taking the product of the last $m+1$ columns of $\MatV$ with $\Q{2^0,2^1,\dots,2^{m}}^T$.  Finally,  use
equation~(\ref{eq:tent-lzc-2}) to obtain $\tyx{Y}{X}$.

In the step 7 we  must generate the surrogate data.  Choose a number $K$ of surrogate  data sets to be generated and
build the  surrogate data matrices $\MatV_k=\Q{\Yvs{t-\tau},\Xv{t-\tau},X_t}$  for $k=1,2,\dots,K$.  Here,  for each
$k$, the sample $\W{\Yvs{t-\tau}}$ is set by redrawing with repetition from the collection $\W{\Yv{t-\tau}}$.  Then,
the surrogate  \acrlong{ter} $\tyxs{Y}{X}$ is obtained  as the  mean value  of $h_k\E{\Yvs{t-\tau},\Xv{t-\tau},X_t}$
over all surrogate realizations.

To estimate $\tyx{X}{Y}$ (transfer entropy from $X$ to $Y$)  and its surrogate $\tyxs{X}{Y}$ just set $y_{t}$ as the
target series,  $x_{t}$  as the source  and repeat the procedure  described in the  steps 3-8.  Finally,  the global
\gls{ter} must be estimated using equation~(\ref{eq:global_te}).
\begin{algorithm}[hb!]
  \caption[caption]{\label{alg:LZTE_1}LeZTER Algorithm.\\\hspace{\textwidth}
                    MATLAB code: \href{https://bitbucket.org/jrinckoar/tentropyrate-lzc/src/master/}
                    {https://bitbucket.org/jrinckoar/tentropyrate-lzc/src/master/}
  }
   \begin{algorithmic}[1]
     \State Binarize the temporal series $x_{t}$ and $y_{t}$ using each median value.
     \State Set $x_{t}$ as target/driven series and $y_{t}$ as source/driver series.
      \algrule
     \State Given a value of $m$  and $\tau$, set the matrix of embedding vectors:
        \begin{equation*}
          \MatV=\Q{\Yv{t-\tau},\Xv{t-\tau},X_t},
        \end{equation*}
        and obtain the sequence $z_n$ (see Fig.~\ref{fig:diagram}).
     \State  Calculate  the  \gls{lzc}  of  $z_n$  and  the  entropy  rate  $h\E{\Yv{t-\tau},\Xv{t-\tau},X_t}$ using
     equation~(\ref{eq:lz-comp}).
     \State Calculate the entropy rate  $h\E{\Xv{t-\tau},X_t}$.  Obtain the corresponding $z_n$ sequence considering
     the submatrix of $\MatV$.
     \State Calculate the \acrlong{ter} $\tyx{Y}{X}$ using equation~(\ref{eq:tent-lzc-2}).
     \State Set a number $K$ of surrogate data sets. For $k=1,2,\dots,K$ form the matrices:
     \begin{equation*}
       \MatV_k=\Q{\Yvs{t-\tau},\Xv{t-\tau},X_t}.
     \end{equation*}
     Calculate   $h_k\E{\Yvs{t-\tau},\Xv{t-\tau},X_t}$    using   equation~(\ref{eq:lz-comp})    and   $\tyxs{Y}{X}$
     using equation~(\ref{eq:tent-lzc-surrogate}).
      \algrule
     \State  Set $y_{t}$  as target  series,  $x_{t}$ as  source  series  and  repeat  the  steps  3-7  to calculate
     $\tyx{X}{Y}$ and $\tyxs{X}{Y}$.
      \algrule
     \State Obtain the global estimation of \acrlong{ter} $\TotalTER$ using equation~(\ref{eq:global_te}).
   \end{algorithmic}
\end{algorithm}
\section{\label{sec:Res}Results}
\begin{figure*}[ht!]
\begin{center}
  \subfloat[\label{fig:H3}]{\includegraphics[width=0.33\textwidth,keepaspectratio]{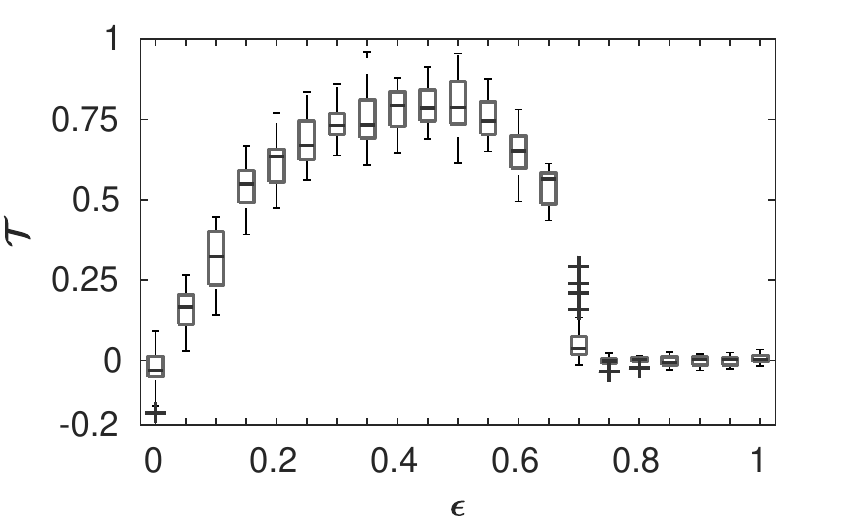}}
  \subfloat[\label{fig:H5}]{\includegraphics[width=0.33\textwidth,keepaspectratio]{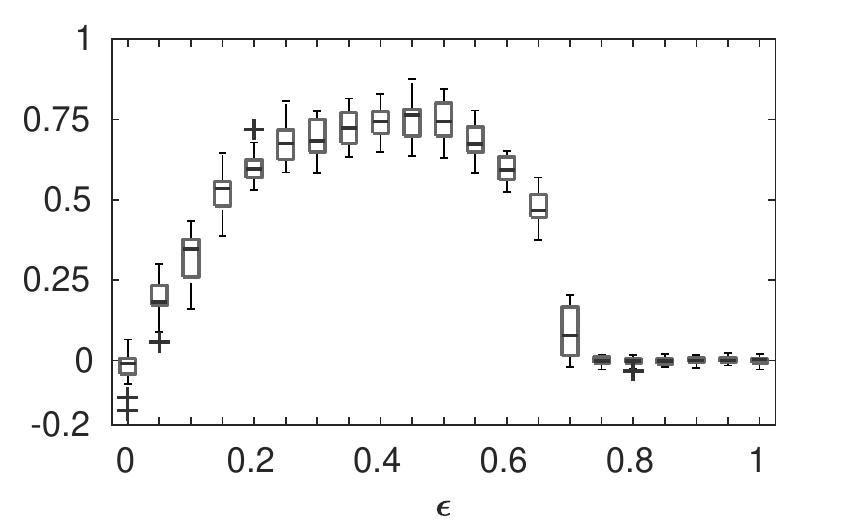}}
  \subfloat[\label{fig:H10}]{\includegraphics[width=0.33\textwidth,keepaspectratio]{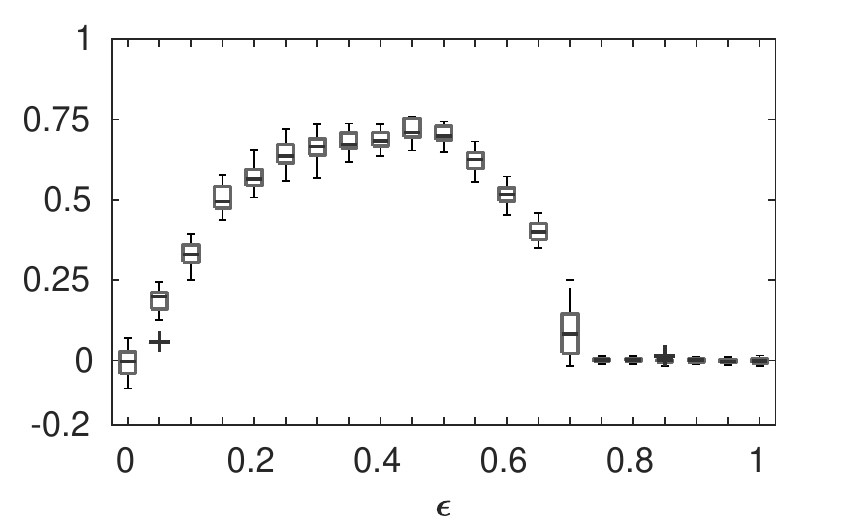}}
\end{center}
\caption{\label{fig:Henon-Henon}Henon-Henon coupled system.  Boxplot of \acrlong{ter}  as a function of the coupling
parameter $\epsilon$.  $\TotalTER$ was calculated with $m=5$ and $\tau=1$ for different data lengths:  (a)~$N=3000$,
(b)~$N=5000$ and (c)~$N=10000$.}
\end{figure*}

\begin{figure*}[ht!]
\begin{center}
  \subfloat[\label{fig:L3}]{\includegraphics[width=0.33\textwidth,keepaspectratio]{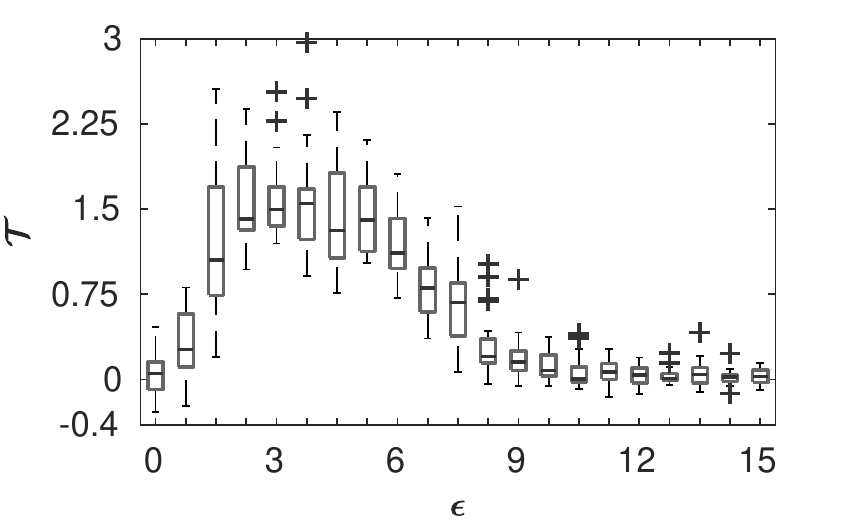}}
  \subfloat[\label{fig:L5}]{\includegraphics[width=0.33\textwidth,keepaspectratio]{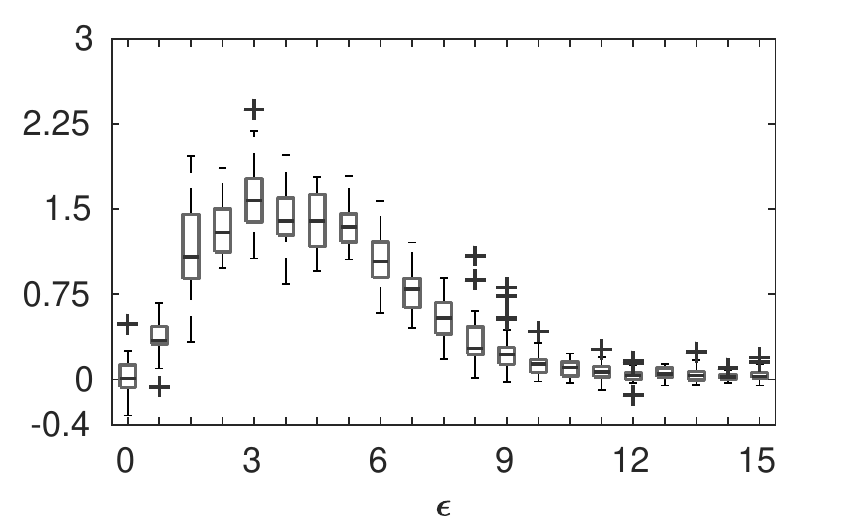}}
  \subfloat[\label{fig:L10}]{\includegraphics[width=0.33\textwidth,keepaspectratio]{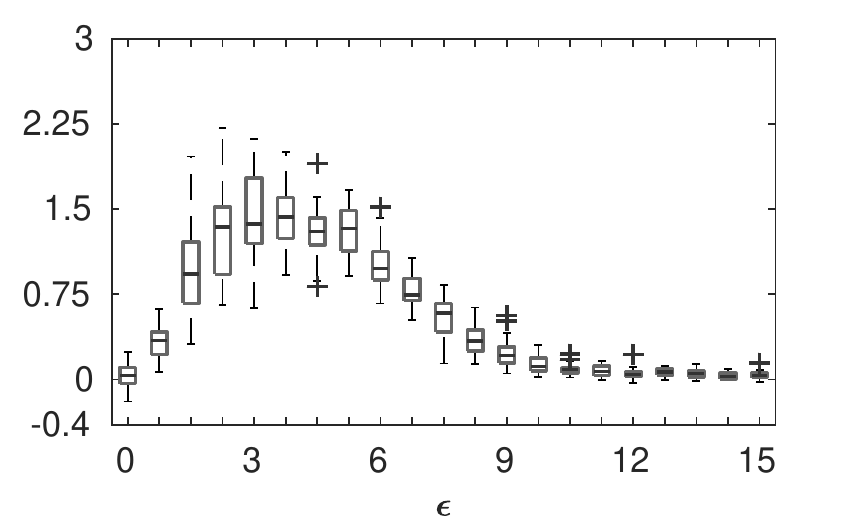}}
\end{center}
\caption{\label{fig:Lorenz-Lorenz}Lorenz-Lorenz  coupled system.  Boxplot  of \acrlong{ter}  as  a  function  of the
coupling  parameter $\epsilon$.  $\TotalTER$  was calculated  with $m=7$  and $\tau=5$  for different  data lengths:
(a)~$N=3000$, (b)~$N=5000$ and (c)~$N=10000$.}
\end{figure*}

\begin{figure*}[ht!]
\begin{center}
  \subfloat[\label{fig:R3}]{\includegraphics[width=0.33\textwidth,keepaspectratio]{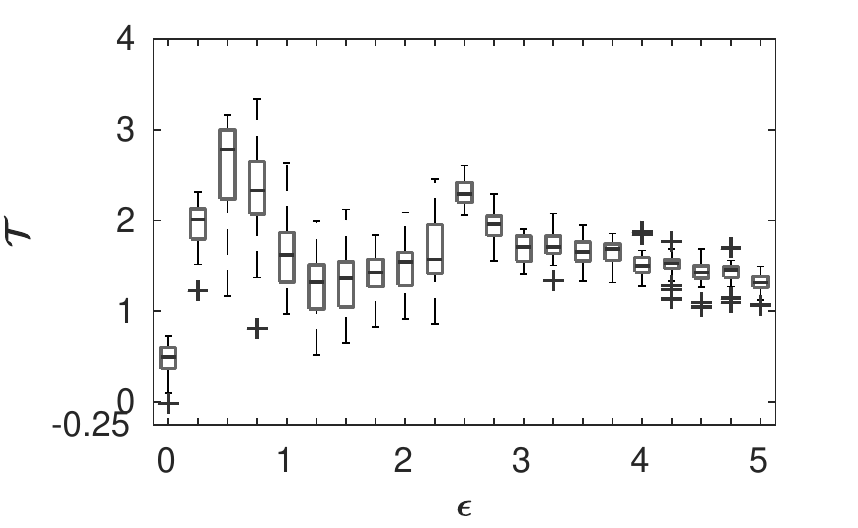}}
  \subfloat[\label{fig:R5}]{\includegraphics[width=0.33\textwidth,keepaspectratio]{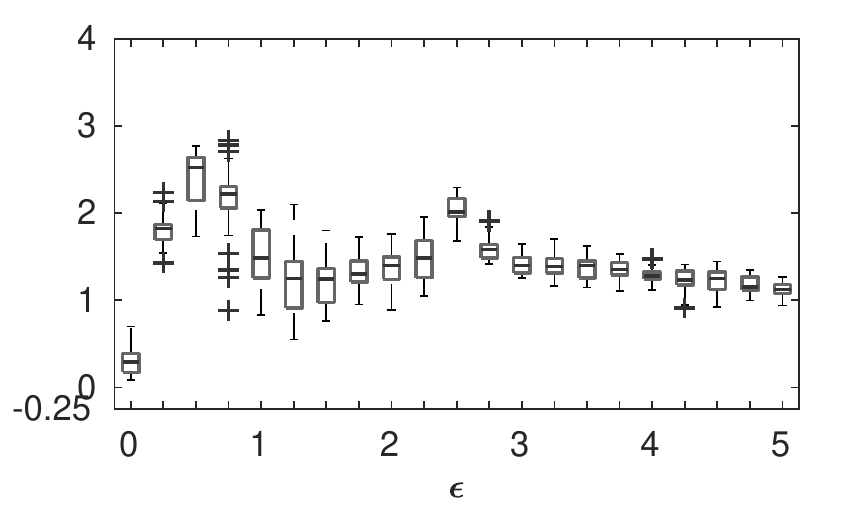}}
  \subfloat[\label{fig:R10}]{\includegraphics[width=0.33\textwidth,keepaspectratio]{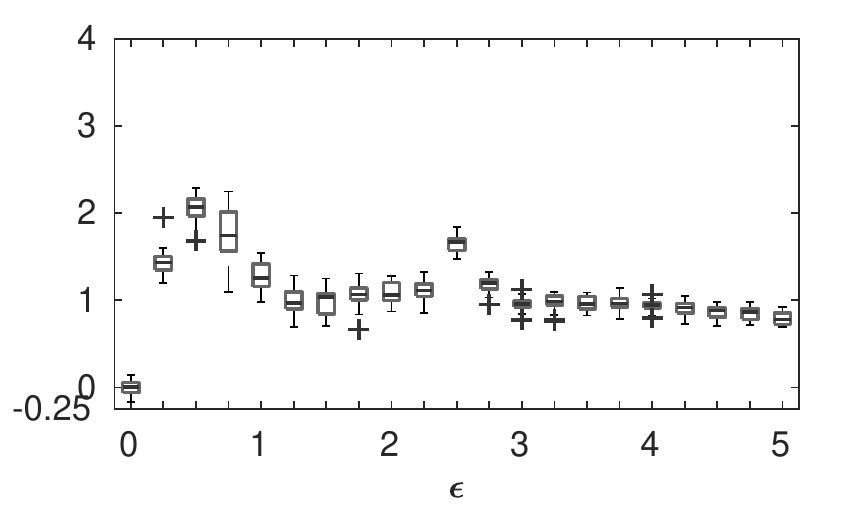}}
\end{center}
\caption{\label{fig:Rosler-Lorenz}Rössler-Lorenz coupled system.  Boxplot  of  \acrlong{ter}  as  a  function of the
coupling parameter  $\epsilon$.  $\TotalTER$ was  calculated with $m=7$  and $\tau=10$  for different  data lengths:
(a)~$N=3000$, (b)~$N=5000$, and (c)~$N=10000$.}
\end{figure*}

We  have  conducted  three  simulations  using  different  unidirectional  coupled  systems:  the  Henon-Henon,  the
Lorenz-Lorenz     and      the     Lorenz     driven      by     Rössler.      The     results      presented     in
Figs.~\ref{fig:Henon-Henon},~\ref{fig:Lorenz-Lorenz}~and~\ref{fig:Rosler-Lorenz}  were  computed   using  values  of
$m$\footnote[2]{The  minimum  embedding  dimension  for the  Henon-Henon  system  is  $m_x  +  m_y=4$  and,  for the
Lorenz-Lorenz  and  Rössler-Lorenz   is  $m_x  +  m_y=6$.}   and  $\tau$  that  met  the   conditions  mentioned  in
Subsection~\ref{sub:implem}.

The first system was the coupled Henon-Henon~\cite{Krakovska2015, Palus2001}:
\begin{equation*}
  \begin{cases}
  y_{1}\Q{n+1} &= 1.4 - y_{1}^{2}\Q{n} -b y_2\Q{n},\\
  y_{2}\Q{n+1} &= y_{1}\Q{n},\\
  x_{1}\Q{n+1} &= 1.4 - \Big(\epsilon y_{1}\Q{n} + \E{1-\epsilon}x_{1}\Q{n}\Big)x_{1}\Q{n} -b x_2\Q{n},\\
  x_{2}\Q{n+1} &= x_{1}\Q{n},\\
  \end{cases}
\end{equation*}
where $b=  0.3$.  For the simulation,  the  coupling parameter $\epsilon$  was varied from  zero to one  in steps of
$0.1$.  For each $\epsilon$,  $200$  realizations were computed using random  initial conditions.  The \acrlong{ter}
was calculated  with $m=\W{2,3,4,5,6,7}$ and $\tau=\W{1,3,5,7,10}$.  This  procedure was  repeated for  data lengths
$N=\W{3000, 5000,10000}$.

The results  are shown in  the Fig.~\ref{fig:Henon-Henon}.  Each plot  presents the global  \gls{ter} ($\TotalTER$),
calculated with  $m=5$ and $\tau=1$,  as  a function of  the coupling parameter  $\epsilon$.  It can be  observed in
Fig.~\ref{fig:H3} ($N=3000$) that  the median value of  estimator $\TotalTER$ is zero for  $\epsilon=0$.  This is an
expected result since  there  is  no  information  flow  between  the  two  systems.  Moreover,  the median value of
$\TotalTER$ increases along with the coupling  parameter until $\epsilon=0.5$.  The positivity of $\TotalTER$ points
out the correct direction of coupling and its  increasing magnitude indicates a rising strength of the coupling.  On
the other hand,  for $\epsilon\geq  0.7$ the median value of $\TotalTER$ is zero.  For  these values of the coupling
parameter  the   Henon-Henon  system  is   synchronized  in  such  a   way  that  both   systems  are  statistically
indistinguishable.  In this  kind of  situations,  $\TotalTER$ is unable  to point  out any  information flow.  This
behaviour has been  already observed on other  Transfer Entropy estimators~\cite{Palus2001,Palus2007,Krakovska2015}.
It can be seen in Figs.~\ref{fig:H5} and~\ref{fig:H10}  ($N=5000$ and $N=10000$,  respectively) that the variance of
$\TotalTER$ decreases as long as the data length is increased.

For the second simulation we have chosen the Lorenz-Lorenz system:
\begin{equation*}
  \begin{cases}
  \dot{y}_{1} & = 10\E{-y_1 + y_2},\\
  \dot{y}_{2} & = \rho_1y_1 -y_2 - y_1y_3,\\
  \dot{y}_{3} & = y_1y_2 - \frac{8}{3}y_3,\\
  \dot{x}_{1} & = 10\E{-x_1 + x_2} + \epsilon\E{y_1-x_1},\\
  \dot{x}_{2} & = \rho_2x_1 -x_2 - x_1x_3,\\
  \dot{x}_{3} & = x_1x_2 - \frac{8}{3}x_3,
  \end{cases}
\end{equation*}
where $\rho_1=28.5$, $\rho_2=27.5$ and $\epsilon\in\W{0,\dots,15}$.  For each value of the coupling parameter, $200$
realization were  computed,  each one starting  from a different  initial condition.  The numerical  integration was
performed using the  ode45 function of Matlab  (algorithm of Dormand and Prince)  with step size $\Delta  t = 0.03$.
For each realization the  first $10000$ data points where discarded.  Then,  $\TotalTER$ was  calculated for all the
combination of  the parameters:  $m=\W{2,3,4,5,6,7}$ and  $\tau=\W{1,3,5,7,10}$.  The above procedure  was applied
varying the data length $N=\W{3000, 5000,10000}$.

In Fig.~\ref{fig:Lorenz-Lorenz} is shown estimator $\TotalTER$ ($m=7$  and $\tau=5$) as a function of $\epsilon$ for
the Lorenz-Lorenz coupled  system.  For $N=3000$ (Fig.~\ref{fig:L3}) it  can be observed that  for uncoupled systems
$\epsilon=0$ the median value of $\TotalTER\approx0$.  As the coupling parameter increases,  $\TotalTER$ is positive
and grows until the synchronization threshold is reached $\epsilon\approx12$~\cite{Krakovska2015}.  From this point,
the value of $\TotalTER$ goes  toward zero despite the systems are coupled.  The same  behavior was displayed by the
symbolic  transfer  entropy~\cite{Staniek2008}  and other  information  flow  estimators  calculated  over  the same
system~\cite{Krakovska2015}.  As well as in the case of the Henon-Henon coupled system,  the variance of $\TotalTER$
decreases as the number of data points is increased (figures~\ref{fig:L5} and~\ref{fig:L10}).

The third system is the Lorenz driven by Rössler (Rössler-Lorenz)  system:
\begin{equation*}
  \begin{cases}
  \dot{y}_{1} & = -\alpha\E{y_2 + y_3},\\
  \dot{y}_{2} & =  \alpha\E{y_1 + 0.2y_2},\\
  \dot{y}_{3} & =  \alpha\Big(0.2 + y_3\E{y_1 - 5.7}\Big),\\
  \dot{x}_{1} & = 10\E{-x_1 + x_2},\\
  \dot{x}_{2} & = 28x_1 - x_2 -x_1x_3 + \epsilon y^{\beta}_2,\\
  \dot{x}_{3} & =  x_1x_2-\frac{8}{3}x_3,
  \end{cases}
\end{equation*}
where $\alpha = 6$, $\beta = 2$ and $\epsilon\in\W{0,0.2,\dots,5}$.  For each value of the coupling parameter, $200$
realization using random initial  conditions were computed.  The numerical integration was  performed using the same
methodology described  for the  Lorenz-Lorenz system,  but  $\Delta t=0.02617$~\cite{Palus2007}.  The  \gls{ter} was
calculated using the same parameter's values of the afore simulations.

In Fig.~\ref{fig:Rosler-Lorenz} the  behavior of $\TotalTER$ as a  function of the coupling parameter  for $m=7$ and
$\tau=10$     is     shown.     For     this     coupled     system     the     synchronization     threshold     is
$\epsilon\approx2.8$~\cite{Krakovska2015,Quiroga2000}.  It can be observed  in Fig.~\ref{fig:R3} ($N=3000$) that the
median value of $\TotalTER$ is always positive, even for $\epsilon=0$.  This means that the $\TotalTER$ estimator is
detecting  false coupling  for $\epsilon=0$.  This  phenomenon has  been  also  observed  in  the  Symbolic Transfer
Entropy~\cite{Staniek2008}.  However,  the  $\TotalTER$  estimator  is  detecting  the  correct  coupling direction.
Figs.~\ref{fig:R5}  and~\ref{fig:R10} display  a similar  behaviour but  notice  that  the  variance  of $\TotalTER$
decreases.
\begin{figure*}[ht!]
\begin{center}
  \subfloat[\label{fig:TentvsM1}]{\includegraphics[width=0.33\textwidth,keepaspectratio]{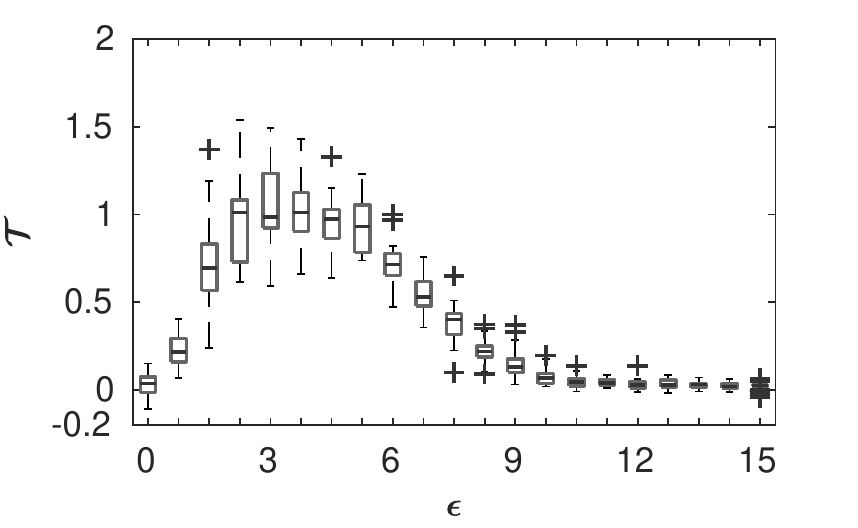}}
  \subfloat[\label{fig:TentvsM2}]{\includegraphics[width=0.33\textwidth,keepaspectratio]{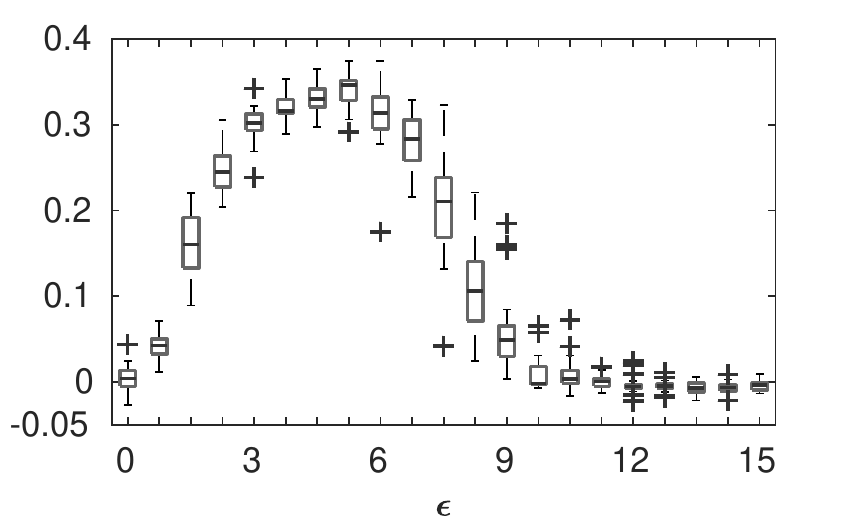}}
  \subfloat[\label{fig:TentvsM3}]{\includegraphics[width=0.33\textwidth,keepaspectratio]{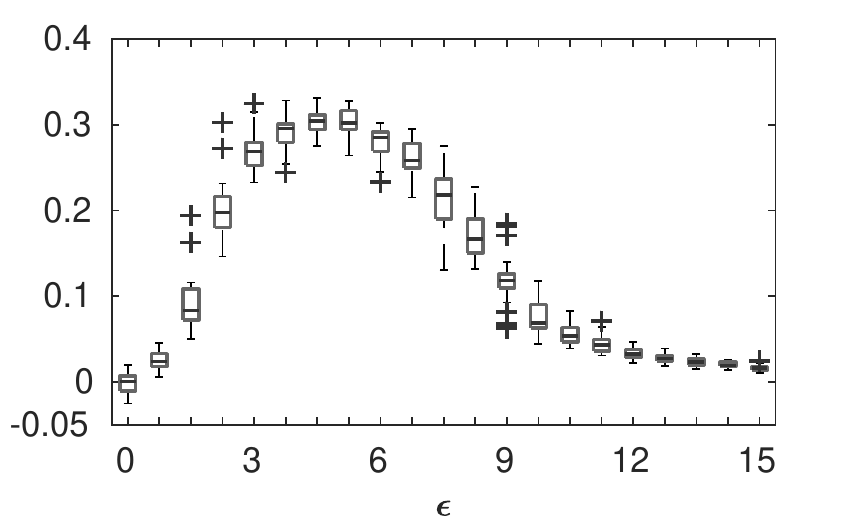}}
\end{center}
\caption{\label{fig:TentvsMET}   Comparison   of   \acrlong{ter}    estimation   with   three   different   methods:
(a)~\acrlong{lzc}  based  method,  (b)~Symbolic  Transfer  Entropy  and  (c)~\glsname{knn}  method.  Boxplot  of the
$\TotalTER$ as a function of the coupling parameter,  calculated for the coupled Lorenz system with $m=5$, $\tau=10$
and $N=10000$.}
\end{figure*}
\begin{figure*}[ht!]
\begin{center}
  \subfloat[\label{fig:TvsN_M1}]{\includegraphics[width=0.33\textwidth,keepaspectratio]{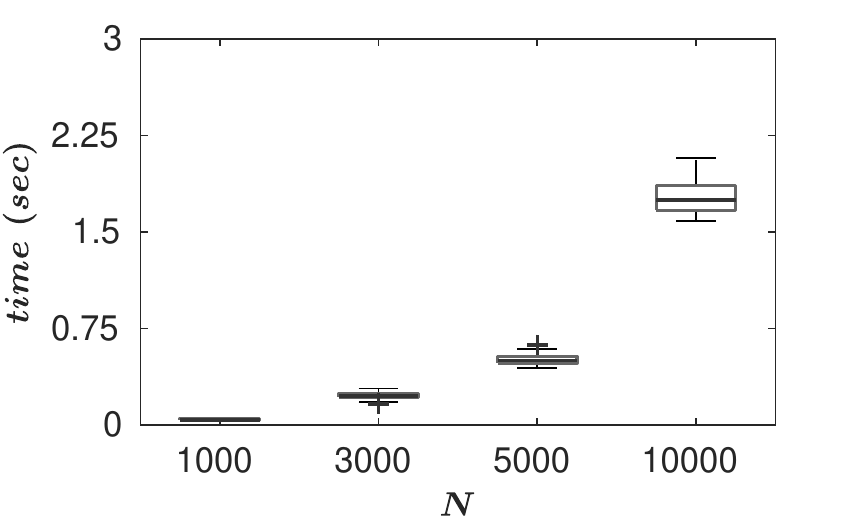}}
  \subfloat[\label{fig:TvsN_M2}]{\includegraphics[width=0.33\textwidth,keepaspectratio]{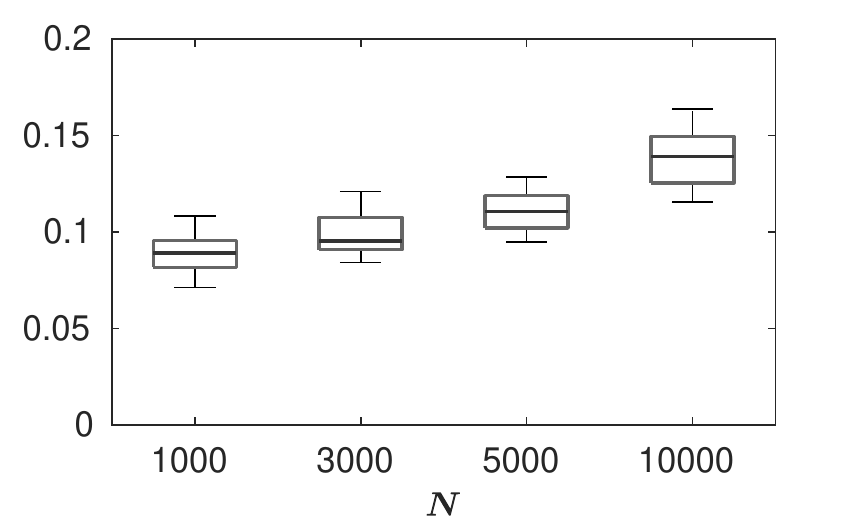}}
  \subfloat[\label{fig:TvsN_M3}]{\includegraphics[width=0.33\textwidth,keepaspectratio]{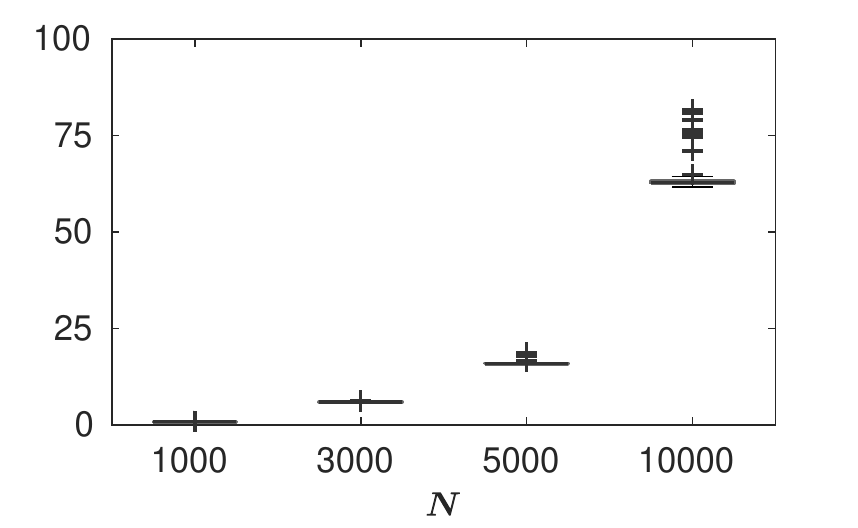}}
\end{center}
\caption{\label{fig:TvsN}  Computation  time  for  a  single as  a  function  of  data  length  ($N$)  for different
\acrlong{ter}   estimation   methods:   (a)~\acrlong{lzc}   based   method,   (b)~Symbolic   Transfer   Entropy  and
(c)~\glsname{knn}  method.  The simulation  was made  using the  coupled  Lorenz  system  with  $m=5$,  $\tau=5$ and
$\epsilon=3$.}
\end{figure*}
\begin{figure*}[ht!]
\begin{center}
  \subfloat[\label{fig:TvsM_M1}]{\includegraphics[width=0.33\textwidth,keepaspectratio]{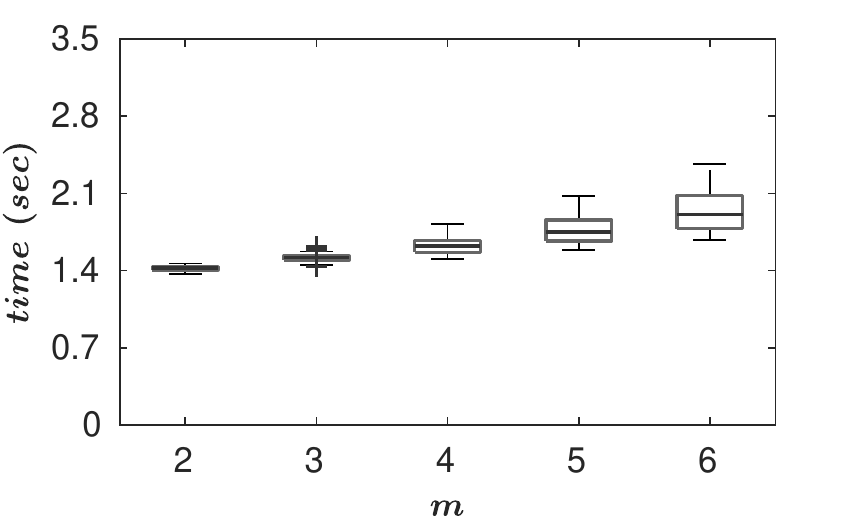}}
  \subfloat[\label{fig:TvsM_M2}]{\includegraphics[width=0.33\textwidth,keepaspectratio]{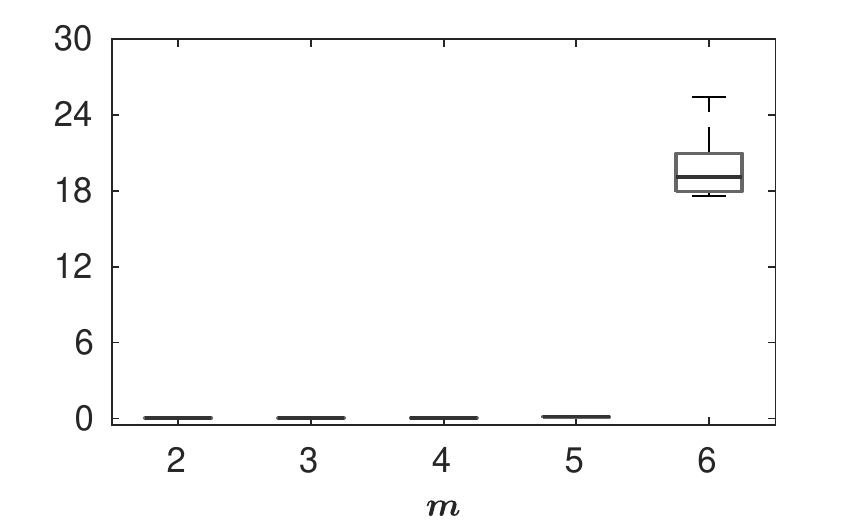}}
  \subfloat[\label{fig:TvsM_M3}]{\includegraphics[width=0.33\textwidth,keepaspectratio]{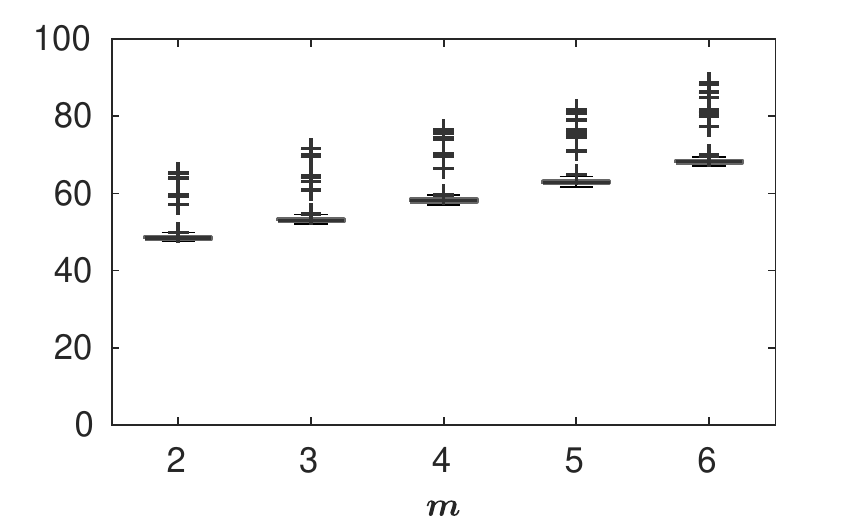}}
\end{center}
\caption{\label{fig:TvsM}  Computation  time  as  a  function   of  the  embedding  dimension  ($m$)  for  different
\acrlong{ter}   estimation   methods:   (a)~\acrlong{lzc}   based   method,   (b)~Symbolic   Transfer   Entropy  and
(c)~\glsname{knn} method.  The simulation  was  made  using  the  coupled  Lorenz  system with parameters $N=10000$,
$\tau=10$ and $\epsilon=3$.}
\end{figure*}
\section{\label{sec:Dis}Discussion}

There are two  methodologies to estimate the \acrlong{ter}  that are similar to our approach.  The  first one is the
Symbolic   Transfer    Entropy~\cite{Staniek2008},    which    finds    its    foundations    in   the   Permutation
Entropy~\cite{Bandt2002}.  The  second  one  can  be found  in~\cite{Lindner2011}  and  is  based  on  the \gls{knn}
estimation method proposed  by Kraskov~\glsname{etal}~\cite{Kraskov2004}.  In order to compare  our methodology with
the  ones mentioned  above,  we have  implemented both  algorithms  and  calculated  the  global  \gls{ter}  and the
computation time of each one of the $200$ realizations of the coupled Lorenz-Lorenz system.  The simulation was done
using $m=\W{2,3,4,5,6}$, $N=\W{1000,  3000, 5000, 10000}$,  $\epsilon=\W{0,\dots,15}$, $\tau=5$ and $K=30$ surrogate
realizations (just for our approach).  The simulation was performed in a cluster with 10 nodes, each one has 2 Intel
Xeon E5-2670 v3 2.5GHz processors of 12 Cores.

In Fig.~\ref{fig:TentvsMET}  we show the global  \gls{ter} as a function  of the coupling parameter  using the three
methods.  It can be observed that $\TotalTER$ presents a very similar behaviour for all the three methods.  However,
it is important to notice that for a given $\epsilon$ their values differ.  This result strength the hypothesis that
our methodology can be used as a information transfer measure.

Results   about   the    comparison    of    computation    times    among    methodologies    can   be   found   in
Figs.~\ref{fig:TvsN}~and~\ref{fig:TvsM},  where we present a boxplot of the computation time of a single realization
as  a  function  of  the  data  lenght  and   the  embedding  dimension,   respectively.   As  it  can  be  seen  in
Figs.~\ref{fig:TvsN_M1},~\ref{fig:TvsN_M2}~and~\ref{fig:TvsN_M3},  the  computation time  of  a  single realization,
increases exponentially  as the data length  increases,  regardless the employed  methodology.  Moreover,  it can be
observed that,  for $m=4$,  the fastest method is the Symbolic Transfer entropy, followed by our methodology and far
away is the \gls{knn} approach.  On the other hand, Figs.~\ref{fig:TvsM_M1},~\ref{fig:TvsM_M2}~and~\ref{fig:TvsM_M3}
show the execution time for a single  realization for different embedding dimensions.  Notice that the computational
cost of each method increases with $m$ in different  ways.  For our approach the increasing is exponential since the
$m$ parameter is linked to the alphabet size in  the Lempel-Ziv's algorithm.  We must point out that our methodology
outperforms   the   Symbolic   Transfer   entropy   approach   for   $m>5$   (compare   the   computation   time  in
Figs~\ref{fig:TvsM_M1}~and~\ref{fig:TvsM_M2} for  $m=6$).  This is  because the computational  cost of  the Symbolic
Transfer entropy increases  with $m$ in a factorial  way.  This suggests that our methodology  has an advantage over
the Symbolic  Transfer entropy  when the  analysis of  high-dimensional systems  is needed.  Finally,  note  for the
greatest embedding dimension here studied ($m=8$) and the longest data length ($N=10000$) tested in this simulation,
our approach performs a single \acrlong{ter} estimation in less than three seconds.

Based on the results,  we can conclude  that the estimator $\TotalTER$ here proposed (equation~(\ref{eq:global_te}))
is able  to detect the  direction and strength  of the information  flow between two  coupled ergodic systems.  This
methodology,  based in the \gls{lzc},  is  computationally fast and it does not assume  any model for the data.  Our
results are  comparable with those  obtained by the Symbolic  Transfer Entropy~\cite{Staniek2008} and  with the ones
reported by Krakovská~\glsname{etal} in~\cite{Krakovska2015}.

The \gls{ter} depends on  two  parameters:  $m$  (the  history  length)  and  $\tau$ (the lag).  For simplicity,  we
considered that these  parameters should be the  same for the embedding  of both time series,  although  they can be
different~\cite{Bossomaier2016}.  We have  observed that the  best results are achieved  when the values  of $m$ and
$\tau$ ensure good reconstruction of the state space.

In future studies,  we will  address the implementation of our methodology using  different embedding parameters for
$x_t$ and $y_{t}$ as well as a different alphabet size.

\section{\label{sec:Con}Conclusions}

In this article we have presented a new methodology  to calculate the \acrlong{ter} between two systems based on the
Lempel-Ziv's complexity.  Because of the  properties  of  the  Lempel-Ziv's  algorithm,  we  were  able to propose a
computationally fast methodology to  estimate the information flow between two  systems.  This methodology have been
assessed using three  unidirectional coupled systems:  Henon-Henon system,  Lorenz-Lorenz  system and Rössler-Lorenz
system.  The results show that our estimator is able to detect the direction and strength of the information flow.

\bibliographystyle{IEEEtran}
\bibliography{bibliography}
\end{document}